\documentclass[prl,twocolumn,showpacs,amsmath,amssymb]
{revtex4}

\usepackage[dvips]{color}
\usepackage{graphicx}

\begin{document}

\title{Emergence of quasi-condensates of hard-core bosons 
at finite momentum}

\author{Marcos Rigol}
\affiliation{Institut f\"ur Theoretische Physik III, Universit\"at 
Stuttgart, Pfaffenwaldring 57, D-70550 Stuttgart, Germany.}

\author{Alejandro Muramatsu}
\affiliation{Institut f\"ur Theoretische Physik III, Universit\"at 
Stuttgart, Pfaffenwaldring 57, D-70550 Stuttgart, Germany.}

\begin{abstract}
An exact treatment of the non-equilibrium dynamics of hard-core bosons 
on one-dimensional lattices shows that, starting from a pure Fock state,
quasi-long-range correlations develop dynamically, and that they lead to 
the formation of quasi-condensates at finite momenta.
Scaling relations characterizing the quasi-condensate and the dynamics 
of its formation are obtained. The relevance of our findings for 
atom lasers with full control of the wave-length by means of a lattice
is discussed.
\end{abstract}

\maketitle

The study of the non-equilibrium dynamics of quantum gases has been 
shown to be a very useful tool for the understanding of their properties 
\cite{dalfovo99}. Since the achievement of Bose-Einstein condensation (BEC) 
\cite{1995}, where the anisotropic expansion of the gas revealed the 
importance of the inter-particle interaction, many experiments studying the 
dynamics of such systems have been developed. Recently, the possibility 
of realizing one-dimensional (1-D) quantum gases has 
attracted a lot of attention. They can be obtained experimentally in very 
elongated traps \cite{1D}, or using optical lattices \cite{greiner01}. 
In these quasi-1D systems it is possible to go from the weakly interacting 
regime to an impenetrable gas of bosons, i.e.\ hard-core bosons (HCB) 
\cite{HCB}. Theoretical studies have analyzed the density evolution 
between these two extreme regimes \cite{santos02}, showing that in general
it does not follow a self-similar solution. In addition, in the HCB limit 
the Fermi-Bose mapping \cite{girardeau60} was generalized to the 
time-dependent case, where the density dynamics revealed dark soliton 
structures, breakdown of the time-dependent mean-field theory, and interference 
patterns of the thermal gas on a ring \cite{girardeau02}.

In 1D systems in equilibrium it was shown that 
quasi-condensates of HCB develop in the homogeneous \cite{homog}, 
harmonically trapped \cite{trap}, and in general cases in 
the presence of a lattice \cite{rigol04_1}. This is because 
the ground state of these systems exhibits off-diagonal 
quasi-long-range order determined by a power-law decay 
of the one-particle density matrix (OPDM) 
$\rho_{ij}\sim |x_i-x_j|^{-1/2}$ \cite{homog,rigol04_1}, 
i.e. there is no BEC in the thermodynamic limit \cite{yang62}.
The occupation of the lowest natural orbital (NO) 
(the highest occupied one) is then proportional to 
$\sqrt{N_b}$ ($N_b$ being the number of HCB) \cite{homog,trap,rigol04_1}. 
The NO ($\phi^\eta_i$) are effective single particle states 
defined as the eigenstates of the OPDM \cite{penrose56}, 
i.e.\ they are obtained by the equation 
$\sum_j \rho_{ij}\phi^\eta_j= \lambda_{\eta}\phi^\eta_i$, 
with $\lambda_{\eta}$ being their occupations.

In the present work we study within an {\it exact} approach the 
non-equilibrium dynamics of HCB in 1D configurations with 
an underlying lattice. The presence of the lattice enhances 
correlations between particles, and its experimental realization 
(the so-called optical lattice) allowed the study of the Mott 
superfluid-insulator transition for soft-core bosons \cite{greiner02_1}. 
In contrast to the continuous case, in a lattice it is possible to create 
pure Fock states of HCB (a HCB per lattice site) where there is no 
coherence in the system. We show that quasi-long-range correlations 
develop in the equal-time-one-particle density matrix (ETOPDM) when 
such states are allowed to evolve freely, and that they lead to the 
formation of quasi-condensates of HCB at finite momentum. In addition 
we obtain an universal power law describing the population of 
the quasi-condensate as a function of time, independent of the initial 
number of particles in the Fock-state. Finally,
we analyze how such systems can be used to create atom lasers with a  
wave-length that can be controlled through the lattice parameter. Our 
analysis is based on a generalization of the method used to study 
the ground-state properties of 1D HCB on a lattice \cite{rigol04_1}.

The HCB Hamiltonian on a lattice can be written as
\begin{equation}
\label{HamHCB} H_{HCB} = -t \sum_{i} \left( b^\dagger_{i} b^{}_{i+1}
+ h.c. \right) + V_\alpha \sum_{i} x_i^\alpha\ n_{i },
\end{equation}
with the addition of the on-site constraints for the creation 
($b^{\dagger}_{i}$) and annihilation ($b_{i}$) operators:
$b^{\dagger 2}_{i}= b^2_{i}=0$, and $\left\lbrace  
b^{}_{i},b^{\dagger}_{i}\right\rbrace =1$. The hopping parameter is denoted 
by $t$ and $n_{i }= b^{\dagger}_{i}b^{}_{i}$ is the particle number operator.
The last term in Eq.\ (\ref{HamHCB}) describes an arbitrary confining 
potential. 

The Jordan-Wigner transformation \cite{jordan28},
\begin{equation}
\label{JordanWigner} b^{\dag}_i=f^{\dag}_i
\prod^{i-1}_{\beta=1}e^{-i\pi f^{\dag}_{\beta}f^{}_{\beta}},\ \ 
b_i=\prod^{i-1}_{\beta=1} e^{i\pi f^{\dag}_{\beta}f^{}_{\beta}}
f_i \ ,
\end{equation}
is used to map the HCB Hamiltonian into the one of non-interacting 
fermions 
\begin{equation}
H_F =-t \sum_{i} \left( f^\dagger_{i}
f^{}_{i+1} + h.c. \right)+ V_\alpha \sum_{i} x_i^\alpha \
n^f_{i }, \label{eq5}
\end{equation}
where $f^\dagger_{i}$ and $f^{}_{i}$ are the creation 
and annihilation operators for spinless fermions, and 
$n^f_{i }=f^\dagger_{i}f^{}_{i}$.

This mapping allows to express the equal-time Green's function 
for the HCB in a non-equilibrium system as
\begin{eqnarray}
\label{green1} G_{ij}(\tau)&=&\langle \Psi_{HCB}(\tau)|
b^{}_{i}b^\dagger_{j}|\Psi_{HCB}(\tau)\rangle \\
&=&\langle \Psi_{F}(\tau)| \prod^{i-1}_{\beta=1}
e^{i\pi f^{\dag}_{\beta}f^{}_{\beta}} f^{}_i f^{\dag}_j
\prod^{j-1}_{\gamma=1} e^{-i\pi f^{\dag}_{\gamma}f^{}_{\gamma}}
|\Psi_{F}(\tau)\rangle, \nonumber
\end{eqnarray}
where $\tau$ is the real time variable, $|\Psi^{G}_{HCB}(\tau)\rangle$ 
is the time evolving wave-function for the HCB and 
$|\Psi^{G}_{F}(\tau)\rangle$ is the corresponding one 
for the non-interacting fermions.

In what follows, we study the time evolution of initial Fock states of HCB 
once they are allowed to evolve freely. Experimentally such states can be 
created by a strong confining potential $V_\alpha(N_ba/2)^{\alpha}\gg t$ 
($a$ is the lattice constant), which avoid vacancies. Since the 
equivalent fermionic system is a non-interacting one, the time evolution of 
an initial wave-function ($|\Psi^I_{F}\rangle$) can be easily calculated
\begin{equation}
\label{time} 
|\Psi_{F}(\tau)\rangle=e^{-iH_F\tau/\hbar}|\Psi^I_{F}\rangle 
= \prod^{N_f}_{\delta=1}\ \sum^N_{\sigma=1} \ P_{\sigma 
\delta}(\tau)f^{\dag}_{\sigma}\ |0 \rangle,
\end{equation} 
which is a product of single particle states, with $N_f$ being 
the number of fermions ($N_f=N_b$), $N$ the number of lattice sites and 
${\bf P(\tau)}$ the matrix of components of $|\Psi_{F}(\tau)\rangle$. 
From here on the method is identical to the one used in Ref.\ 
\cite{rigol04_1}, the action of 
$\prod^{j-1}_{\gamma=1} e^{-i\pi f^{\dag}_{\gamma}f_{\gamma}}$ 
on the right in Eq.\ (\ref{green1}) generates only a 
change of sign on the elements $P_{\sigma \delta}(\tau)$ 
for $\sigma \leq j-1$, and the further creation of a particle at site 
$j$ implies the addition of one column to ${\bf P(\tau)}$ with the 
element $P_{jN_f+1}(\tau)=1$ and all the others equal to zero 
(the same applies to the action of $\prod^{i-1}_{\beta=1} e^{i\pi 
f^{\dag}_{\beta}f_{\beta}} f_i$ on the left of Eq.\ (\ref{green1})). 
Then the HCB Green's function can be calculated exactly as \cite{rigol04_1}
\begin{eqnarray}
\label{determ}
G_{ij}(\tau)
&=&\langle 0 | \prod^{N_f+1}_{\delta=1}\ \sum^N_{\beta=1} \ 
P'^{A}_{\beta \delta}(\tau)f_{\beta} 
\prod^{N_f+1}_{\sigma=1}\ \sum^N_{\gamma=1} \ P'^{B}_{\gamma 
\sigma}(\tau)f^{\dag}_{\gamma}\ |0 \rangle \nonumber \\
&=&\det\left[ \left( {\bf P}^{'A}(\tau)
\right)^{\dag}{\bf P}^{'B}(\tau)\right],
\end{eqnarray}
where ${\bf P'}^{A}(\tau)$ and ${\bf P'}^{B}(\tau)$ are the new matrices
obtained from ${\bf P}(\tau)$ when the required signs are changed and
the new columns added. The evaluation of $G_{ij}(\tau)$ is done 
numerically, and the ETOPDM ($\rho_{ij}(\tau)$) is determined by the expression 
$\rho_{ij}(\tau)=\left\langle b^\dagger_{i}b_{j}\right\rangle_\tau 
=G_{ji}(\tau)+\delta_{ij}\left(1-2 G_{ii}(\tau)\right)$.

\begin{figure}[h]
\includegraphics[width=0.48\textwidth,height=0.225\textwidth]
{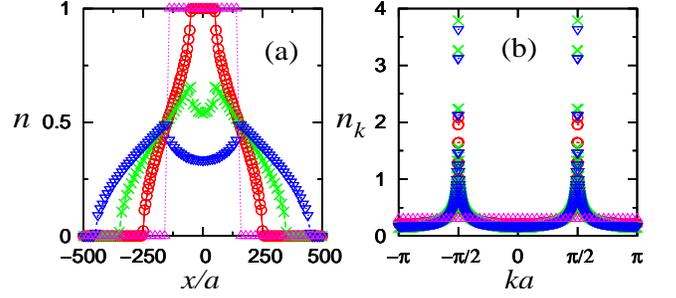}
\caption{(color online). Evolution of density (a) and momentum (b) 
profiles of 300 HCB in 1000 lattice sites. The times are 
$\tau=0$ (\textcolor{magenta}{$\triangle$}), 
$50\hbar/t$ (\textcolor{red}{$\bigcirc$}), 
$100\hbar/t$ (\textcolor{green}{$\times$}), and 
$150\hbar/t$ (\textcolor{blue}{$\nabla$}). Positions (a)
and momenta (b) are normalized by the lattice constant $a$.}
\label{Perfil}
\end{figure}
Figure \ref{Perfil} shows density profiles (a), and their corresponding  
momentum distribution functions (MDF) (b) for the time evolution of an 
initial Fock state. Initially, the MDF 
is flat as corresponds to a pure Fock state, and during the evolution 
of the system sharp peaks appear at $k=\pm \pi/2a$. 
Notice that although in the equivalent fermionic system
the density profiles are equal to the ones of the HCB, 
the MDF remains flat since the fermions do not interact, making evident the 
non-trivial differences in the off-diagonal correlations between both systems.

\begin{figure}[h]
\includegraphics[width=0.48\textwidth,height=0.235\textwidth]
{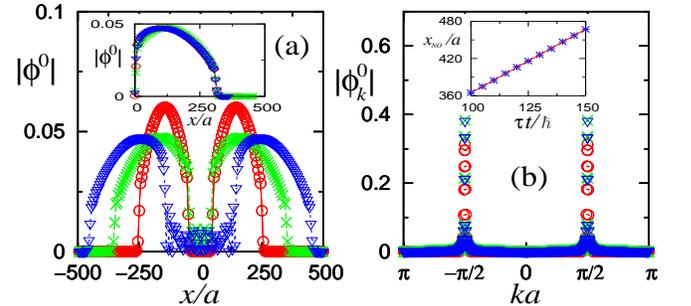}
\caption{(color online). Lowest NO evolution (a) and its fourier 
transform (b). The times are $\tau=50\hbar/t$ (\textcolor{red}{$\bigcirc$}), 
$100\hbar/t$ (\textcolor{green}{$\times$}), and 
$150\hbar/t$ (\textcolor{blue}{$\nabla$}). The insets show: (a)
the superposed right lobe of the lowest NO at 
$\tau=100\hbar/t$ (\textcolor{green}{$\times$}), 
$125\hbar/t$ (\textcolor{red}{$\bigcirc$}), 
and $150\hbar/t$ (\textcolor{blue}{$\nabla$}); (b) the evolution of 
the lowest NO right lobe position, the line has a slope 2.}
\label{NO}
\end{figure}

Since the peaks in the MDF may correspond to quasi-condensates 
at finite momenta, we diagonalize the ETOPDM to study the NO. 
In Fig.\ \ref{NO}(a) we show the lowest NO 
($|\phi^0|$ since $\phi^0$ is complex) corresponding to the results in 
Fig.\ \ref{Perfil} for $\tau>0$ (at $\tau=0$ the NO are delta functions 
at the occupied sites). They are two-fold degenerate, corresponding to 
right- (for $x>0$) and left- (for $x<0$) moving solutions. 
It can be seen that some time after the $n=1$ 
plateau disappears from the system, the NO lobes almost do not change 
their form and size (see inset in Fig.\ \ref{NO}(a) for the 
right lobe at three different times). Furthermore, they move with constant 
velocities $v_{NO}=\pm 2at/\hbar$ (see inset in Fig.\ \ref{NO}(b) for the 
positive one). These are in fact, the group velocities 
$v=1/\hbar\ \partial \epsilon_k/\partial k$
for a dispersion $\epsilon_k=-2t\cos ka$ (the one of HCB on a lattice)
at $k=\pm \pi/2a$. This is further confirmed by considering 
the Fourier transforms of the lowest NO Fig.\ \ref{NO}(b), which show 
sharp peaks centered at quasi-momenta $k=\pm \pi/2a$. 
The peak at $k=+\pi/2a$ appears due to the 
Fourier transform of the right lobe in Fig.\ \ref{NO} 
(a) (i.e.\ $\phi^0_i \approx |\phi^0_i| e^{i\pi x_i/2a}$ for $x_i>0$), 
and the one at $k=-\pi/2a$ due to the fourier transform 
of the left lobe (i.e.\ $\phi^0_i \approx |\phi^0_i| e^{-i\pi x_i/2a}$ 
for $x_i<0$). Studying the Fourier transform of the other NO we have 
seen that they have very small or zero weight at $k=\pm \pi/2a$ so that 
we can conclude that the peaks at $n_{k=\pm \pi/2a}$ are reflecting 
the large occupation of the lowest NO. 

\begin{figure}[h]
\includegraphics[width=0.48\textwidth,height=0.23\textwidth]
{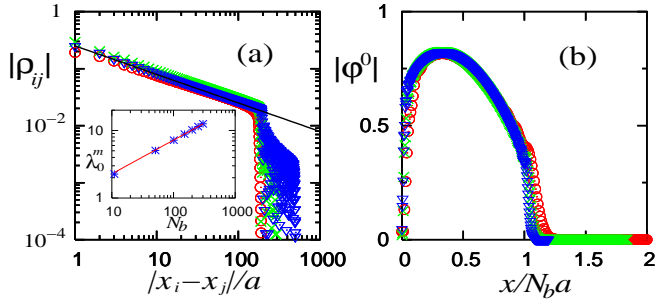}
\caption{(color online). (a) ETOPDM for: 
$\tau$=$50\hbar/t$ (\textcolor{red}{$\bigcirc$}), 
$100\hbar/t$ (\textcolor{green}{$\times$}), and 
$150\hbar/t$ (\textcolor{blue}{$\nabla$}), the line is 
$0.25\ |(x_i-x_j)/a|^{-1/2}$ (see text for details). 
(b) Scaled lowest NO right lobe's {\it vs} $x/N_ba$ for 
$N_b=101$ (\textcolor{red}{$\bigcirc$}), 
$201$ (\textcolor{green}{$\times$}), and 
$301$ (\textcolor{blue}{$\nabla$}). 
The inset in (a) shows the maximum occupation of the lowest NO 
{\it vs} $N_b$ of the initial Fock state, the line is 
$0.72\ N_b^{1/2}$.}
\label{ETOPDM}
\end{figure}

Whether the large occupation of the lowest NO corresponds to a 
quasi-condensate, can be determined studying the ETOPDM. Fig.\ \ref{ETOPDM}(a) 
shows the results for $|\rho_{ij}|$ (with $i$ taken at the beginning 
of the lowest NO left lobe's and $j>i$) at the same times of 
Figs.\ \ref{Perfil}, \ref{NO}. It can be seen that off-diagonal 
quasi-long-range order develops in the ETOPDM. It is reflected by a 
power-law decay of the form $|\rho_{ij}|=0.25\ |(x_i-x_j)/a|^{-1/2}$, that 
remains almost unchanged during the evolution of the system. 
A careful inspection shows that this power-law behavior is restricted 
to the regions where each lobe of the lowest NO exists, outside these 
regions the ETOPDM decays faster. This quasi-coherent behavior in two 
different segments of the system is the reason for the 
degeneracy found in the lowest NO. In addition, a detailed analysis 
of the ETOPDM Fourier's transform shows that the peak in the MDF at 
$k=+\pi/2a$ is originated by components of the ETOPDM with $x_i,x_j>0$, and 
the one at $k=-\pi/2a$ by the components with $x_i,x_j<0$, so that in 
the regions of the lobes 
$\rho_{ij}\approx 0.25\ |(x_i-x_j)/a|^{-1/2}e^{i\pi(x_i-x_j)/2a}$ 
for $x_i,x_j>0$, and 
$\rho_{ij}\approx 0.25\ |(x_i-x_j)/a|^{-1/2}e^{-i\pi(x_i-x_j)/2a}$ 
for $x_i,x_j<0$. The prefactor of the power law (0.25) was found to be 
independent of the number of particles in the initial 
Fock state.

After having observed off-diagonal quasi-long-range correlations in 
the ETOPDM, it remains to consider how the occupation of the lowest NO
behaves in the thermodynamic limit, in order to see whether it corresponds 
to a quasi-condensate. The only new information we need at this point is the 
scaling relation (if any) of the modulus of the lowest NO 
with the number of particles in the initial Fock state. 
For this, we consider the NO at those times, where as shown in 
the inset of Fig.\ \ref{NO}, they almost do not change in time,
i.e.\ after the plateau with $n=1$ disappears.
In Fig.\ \ref{ETOPDM}(b) we show that a scaled NO 
($|\varphi_0|=N_b^{1/2}|\phi^0|$) exists when its size is normalized 
by $N_ba$. (The results for the left lobe are identical due to inversion 
symmetry.) Figure \ref{ETOPDM}(b) also shows that the lobe size $L$ is 
approximately $N_ba$. Then evaluating 
$\lambda_0=\sum_{ij}\phi^{*0}_i \rho_{ij}\phi^0_j$ 
as the double of the integral over a single lobe ($L\gg a$) one obtains
\begin{eqnarray}
\lambda_0 &\sim& 2/a^2
\int^L_{0}dx \int^L_{0}dy \ \phi^{*0}(x)\rho(x,y) \phi^0(y)
\\ &=& N_b^{1/2}
\int^{1}_{0}dX \int^{1}_{0}dY 
\frac{|\varphi^0(X)|0.25|\varphi^0(Y)|}{|X-Y|^{-1/2}} 
=A\sqrt{N_b}, \nonumber \label{lambda0}
\end{eqnarray}
where we did the change of variables $x$=$XN_ba$, $y$=$YN_ba$, 
$\phi^0$=$N_b^{-1/2}\varphi^0$, and we notice that the phase 
factors between the NO and the ETOPDM cancel out. The integral over $X,Y$ 
is a constant that we call $A$. A confirmation of the validity of the 
previous calculation is shown in the inset of Fig.\ \ref{ETOPDM}(a). 
There we plot the maximum occupation of the lowest NO (reached when the 
lobes of the NO have a stable form), as a function of the number of 
particles in the initial Fock state. The $\sqrt{N_b}$ power-law behavior 
is evident and a fit allowed to obtain the constant $A=0.72$.
Hence, this and the previous results demonstrate that the peaks of the MDF 
in Fig.\ \ref{Perfil}(b) correspond to quasi-condensates with
momenta $k=\pm \pi/2a$.

The appearance of such quasi-condensates at $k=\pm \pi/2a$ 
can be understood on the basis of total energy conservation.
Given the dispersion relation of HCB on a lattice,  
since the initial Fock state has a flat MDF, its total energy
is $E_T = 0$. Would all the particles condense into one state,
it would be the one with an energy corresponding to 
$\overline{\epsilon}_k = E_T/N$.
Taking into account the dispersion relation for HCB on a lattice
($\epsilon_k=-2t\cos ka$), $\overline{\epsilon}_k=0$ corresponds
to $k=\pm \pi/2a$. Actually, in the 1D case there is only
quasi-condensation, so that the argument above applies only to 
maximize the occupation of a given state. In addition, the minimum 
in the density of states at these quasi-momenta reinforces 
quasi-condensation into a single momentum state.

\begin{figure}[h]
\includegraphics[width=0.48\textwidth,height=0.235\textwidth]
{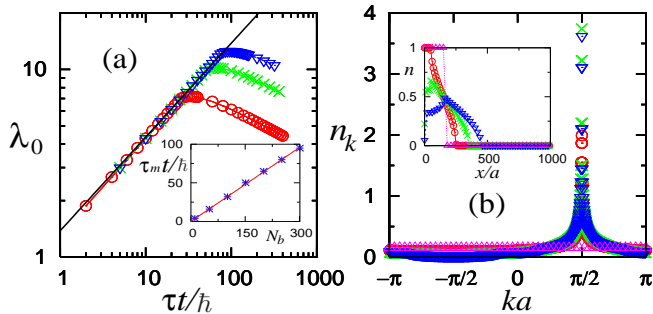}
\caption{(color online). (a) Time evolution of the lowest NO 
occupation for $N_b=101$ (\textcolor{red}{$\bigcirc$}), 
$201$ (\textcolor{green}{$\times$}), and 
$301$ (\textcolor{blue}{$\nabla$}), the straight line 
is $1.38\sqrt{\tau t/\hbar}$. The inset shows the 
time at which the maximum occupation of the NO is reached 
($\tau_m t/\hbar$) {\it vs} $N_b$, the line is $0.32N_b$. 
(b) Density profiles (inset) and MDF of 150 HCB evolving only 
to the right in 1000 lattice sites for 
$\tau=0$ (\textcolor{magenta}{$\triangle$}), 
$50\hbar/t$ (\textcolor{red}{$\bigcirc$}), 
$100\hbar/t$ (\textcolor{green}{$\times$}), and 
$150\hbar/t$ (\textcolor{blue}{$\nabla$}).}
\label{LASER}
\end{figure}

The process of formation of the quasi-condensate is also characterized
by a power law. In Fig.\ \ref{LASER}(a) we plot
the occupation of the lowest NO as a function of the evolution 
time. The {\it log-log} scale shows that the occupation of the 
quasi-condensate increases quickly and decays slowly. A more
detailed examination shows that the population of the quasi-condensate
increases in an universal way ($1.38\sqrt{\tau t/\hbar}$, continuous line in 
Fig.\ \ref{LASER}(a)) independently of the initial number of particles 
in the Fock state. The power law is determined by the universal behavior 
of the off-diagonal part of the ETOPDM shown before, and the fact that
during the formation of the quasi-condensate, the NO increases its size
linearly with time at a rate given by $|v_{NO}| = 2at/\hbar$. 
Such a power law is followed almost up to the point where the maximal 
occupation is reached. The time at which it is reached depends linearly 
on the number of particles of the initial Fock state 
$\tau_m=0.32N_b\hbar /t$ as shown in the inset of Fig.\ \ref{LASER}(a).
For a tipical experimental setup of rubidium atoms, in a lattice with a
recoil energy of $E_r=20$kHz and a depth of 20$E_r$, the time $\tau_m$ 
can be estimated as $\tau_m\sim 5.7N_b$(ms).

The results above showing that a quasi-coherent matter 
front forms spontaneously from Fock states of HCB, suggest that
such an arrangement could be used to create atom lasers with a wave-length 
that can be fully controlled given the lattice parameter $a$. 
No additional effort is needed to 
separate the quasi-coherent part from the rest since the quasi-condensate is 
traveling at the maximum possible velocity on the lattice so that the front 
part of the expanding cloud is the quasi-coherent part. 
The actual realization would imply to restrict the evolution of the initial 
Fock state to one direction, as shown in Fig.\ \ref{LASER}(b), where we 
display the MDF for 150 HCB restricted to evolve 
to the right in 1000 lattice sites at the same evolution times of Fig.\ 
\ref{Perfil}. It can be seen that the values of $n_{k=\pi/2}(\tau)$ are 
almost the same in both situations, although in Fig.\ \ref{LASER}(b) 
the initial Fock state has almost half of the particles. The same occurs to 
the occupation of the lowest NO that in the latter case is not degenerated 
anymore. 

The previous results suggest how to proceed in order to obtain  
lasers in higher dimensional systems where real condensation can occur 
\cite{lieb02}. One can employ Mott insulator states with one particle per 
lattice site created by a very strong on-site repulsive potential $U$. 
The latter is required in order to obtain a close realization of a 
pure Fock state, since quantum fluctuations of the particle number present 
in a Mott insulator for any finite $U$ \cite{hubbard} will be strongly 
suppressed. Then the geometry of the lattice should be designed in order 
to restrict the evolution of the Mott insulator to one direction only, 
and to have a low density of states around the mean value of energy per 
particle in the initial state. With these conditions the sharp features 
observed in 1D should be reproduced by a condensate in higher dimensions.

In summary, we have studied the non-equilibrium dynamics of a 
Fock states of HCB when they are allowed to evolve freely. We have 
shown that quasi-long-range correlations develop dynamically in these 
systems and that they lead to the formation of quasi-condensates of HCB 
at quasi-momenta $k\pm \pi/2a$. We have studied the dynamics of the formation 
of these quasi-condensates, and their occupations were found to scale 
proportionally to $\sqrt{N_b}$. These systems can be used to create 
atom lasers since the quasi-condensate develops quickly and decays slowly. 
The wave-length of such lasers is simply determined by the lattice parameter. 
Finally, we have discussed the possibility of creating atom lasers in higher 
dimensional systems where true condensation can occur.

We thank HLR-Stuttgart (Project DynMet) for allocation of computer
time, and SFB 382 for financial support. We are gratefull to 
F. de Le\'on for useful discussions.

\end{document}